\documentclass[12pt]{article}

\usepackage{graphicx}
\usepackage{amstext,amsmath,amssymb,amsfonts,bbm,slashed}
\usepackage[latin1]{inputenc}
\usepackage{epsfig}
\usepackage{hyperref}
\usepackage{amsthm}
\usepackage{subfigure}
\usepackage{color}

\makeatletter

\usepackage{color}

\usepackage{dcolumn}% Align table columns on decimal point
\usepackage{bm}% bold math
\usepackage{hyperref}% add hypertext capabilities
\newtheorem{definition}{Definition}

\newcommand{\bea}{\begin{eqnarray}}
\newcommand{\eea}{\end{eqnarray}}
\newcommand{\beq}{\begin{equation}}
\newcommand{\eeq}{\end{equation}}

\newcommand{\br}{\begin{array}}	
\newcommand{\er}{\end{array}}

\newcommand{\mR}{\mathbb{R}}

\begin{document}

%\begin{titlepage}

%\begin{flushright}
%ICMPA-MPA/2013/10
%\end{flushright}

\vspace{20pt}

\begin{center}

\hspace{-0.5cm}
{\Large\bf Position-dependent noncommutative quantum models:
\\ 
\medskip 
Exact solution of the harmonic oscillator
}
\vspace{15pt}

{\large Dine Ousmane Samary }

\vspace{15pt}
{\sl Perimeter Institute for Theoretical Physics\\
31 Caroline St. N.
Waterloo, ON N2L 2Y5, Canada}

\vspace{0.5cm}

{\sl International Chair in Mathematical Physics and Applications\\ (ICMPA-UNESCO Chair), University of Abomey-Calavi,\\
072B.P.50, Cotonou, Rep. of Benin}\\
\vspace{5pt}
E-mail:  {\sl   ousmanesamarydine@yahoo.fr}

\vspace{10pt}

\begin{abstract}
This paper is devoted to find the exact solution of the harmonic oscillator in a position-dependent $4$-dimensional noncommutative phase space. The noncommutative phase space  that we consider is described by the commutation relations between coordinates and momenta: $[\hat{x}^1,\hat{x}^2]=i\theta(1+\omega_2 \hat x^2)$, $[\hat{p}^1,\hat{p}^2]=i\bar\theta$, $[\hat{x}^i,\hat{p}^j]=i\hbar_{eff}\delta^{ij}$. We give an analytical method to solve the  eigenvalue problem of the harmonic oscillator within this deformation algebra.
\end{abstract}

\end{center}

%\noindent  Pacs numbers:   05.45.-a,\,\, 46.25.Cc
%\\
\setcounter{footnote}{0}

%\end{titlepage}

%\maketitle
%\tableofcontents
%%%%%%%%%%%%%%%%%%%%New Section%%%%%%%%%%%%%%%%%%%%
%%%%%%%%%%%%%%%%%%%%%%%%%%%%%%%%%%%%%%%%%%%%%%%
\section{Introduction}
Noncommutative (NC) geometry plays an increasing role
in the search of a unifying theory of gravity and quantum mechanics and is a framework built for   
understanding  physics at short distances. Within this framework,
the past two decades have witnessed important progresses toward the solution of various quantum models, in particular, the harmonic oscillator in 
NC spaces. There exists a  large number of papers which  address this class of problem. We will focus on the most recent developments discussing particular tractable models and specific ways to realize NC spaces called Moyal spaces \cite{Kim:1988mn}-\cite{Jing:2008zzc}.

 The Moyal type NC space is a concrete proposal for a space  where the coordinate operators $\hat x^\mu$ satisfy the commutation relation
\bea\label{com1}
[\hat{x}^\mu,\hat{x}^\nu]=i\theta^{\mu\nu}
\eea
and where $\theta^{\mu\nu}$ is an antisymmetric tensor of space dimension $(length)^2$. 
The noncommutativity specified by (\ref{com1}) can be as well realized in terms of a star product. In this point of view, the ordinary multiplication of functions is replaced 
by the Moyal star product defined for $f,g\in C^\infty(\mR^D)$ by
\beq\label{Moyal}
(f\star g)(x)={\bf m}\Big[\exp\Big(\frac{i}{2}\theta^{\mu\nu}\partial_\mu\otimes\partial_{\nu}
\Big)(f\otimes g)(x)\Big],\quad {\bf m}(f\otimes g)(x)=f(x)\cdot g(x).
\eeq
Then the commutation relation (\ref{com1}) becomes 
 \beq \label{Ben}
[x^\mu,x^\nu]_\star=x^\mu\star x^\nu-x^\nu\star x^\mu=i\theta^{\mu\nu}
\eeq
with now commuting 
 coordinates $x^\mu$.
The noncommutativity of space coordinates can be naturally incorporated into the quantum field theory framework.  Subsequently, NC field
theories and quantum mechanics have studied extensively \cite{Szabo:2001kg}. There is however a more general structure  extending the Moyal brackets \eqref{Ben}.  Consider that one replaces  the commutation relation (\ref{com1})  by   following  \cite{Aschieri:2008zv}-\cite{Hounkonnou:2010yk}
\bea\label{com2}
[\hat{x}^\mu,\hat{x}^\nu]=i\theta^{\mu\nu}  e(\hat x)
\eea
where $e(\hat x)$ is an arbitrary dimensionless function which depends on the coordinates.     The same space can be again realized using another star product called the twisted Moyal product generalizing \eqref{Moyal}.
Taking $e(x)$  positive, the star product 
\bea\label{Moyalt}
(f\star g)(x)={\bf m}\Big[\exp\Big(\frac{i}{2}\theta^{\mu\nu}\sqrt{e(x)}\partial_\mu
\otimes\sqrt{e(x)}\partial_{\nu}
\Big)(f\otimes g)(x)\Big]
\eea
can be used to generate 
\bea\label{twistcom}
[x^\mu,x^\nu]_\star=i\theta^{\mu\nu}e(x)
\eea
now extending \eqref{Ben}.
 The choice of the function $e(x)$ depends on the physical considerations which may   encode minimal length \cite{Kempf:1996nk}  or the  integrability of some dynamical Hamiltonians \cite{Hounkonnou:2011zz}.

 We emphasize the fact that a necessary condition for having an  associative    star product from (\ref{Moyalt}) is given by $\partial_{[\mu}e(x)\partial_{\nu]}f=0,\,\,\forall f\in C^1(\mathbb{R}^D)$ \cite{Aschieri:2008zv}. This    is not  however a sufficient condition.
 The associativity of the twisted star product implies the Jacobi 
identity
\bea
J(\mu,\nu,\rho)=[x^\mu,[x^\nu,x^\rho]_\star]_\star+[x^\rho,[x^\mu,x^\nu]_\star]_\star+[x^\nu,[x^\rho,x^\mu]_\star]_\star=0.
\eea
The particular case of the structure function
\bea
e(x)=1+\omega_\mu x^\mu
\eea
where $\omega_\mu x^\mu$ is dimensionless and $\omega_\mu\in\mR$,
leads to
\bea
J(\mu,\nu,\rho)=-e(x)\omega_\sigma\Big(\theta^{\nu\rho}
\theta^{\mu\sigma}+\theta^{\mu\nu}
\theta^{\rho\sigma}+\theta^{\rho\mu}
\theta^{\nu\sigma}\Big).
\eea
For such a choice of the function $e(x)$, the associativity of the star product \eqref{Moyalt} can be shown even for the non-vanishing tensor $\omega_\sigma$ \cite{Hounkonnou:2011zz}. From this point, the authors of  \cite{Hounkonnou:2011zz} were able to derive the equivalent of the 
so-called matrix basis of the Moyal plane \cite{Grosse:2003nw,deGoursac:2008rb}.

NC spaces can be slightly more general than the above. For instance there are several developments  around the so called NC quantum mechanics  \cite{Diao:2011zz,Wang:2009vm,Geloun:2009gv,Scholtz:2008zu,Rohwer:2010zq}.
NC quantum mechanics \cite{Scholtz:2008zu}   can be also described by  introducing commutation relations between coordinate and momentum even also  between  momentum and momentum. Thus \eqref{twistcom} can be extended to a $2D$  NC phase space as follows
\bea\label{deformation}
[x^1,x^2]_\star=i\theta e(x),\qquad [ x^1, p^1]_\star=i\hbar_{eff},\quad
[x^2, p^2]_\star=i\hbar_{eff},\qquad[ p^1, p^2]=i\bar\theta
 \eea
where   
 $\theta$, $\bar{\theta}$ and $\hbar_{eff}$ are  constant but 
 $e(x)=1+\omega_1 x^1+\omega_2 x^2$ is still a function. The present work  highlights the spectrum of the harmonic oscillator in this twisted NC phase space   defined by  the  commutation relations \eqref{deformation}, with the restriction  $\omega_1=0$. We show, using a particular transformation of the basic degrees of freedom that the total nonlinear harmonic oscillator Hamiltonian factorizes. From that point, the model becomes solvable.

The paper is organized as follows. In section \ref{sec1},\, we give some useful results concerning  the deformation of the NC phase space  (\ref{deformation}). Then, the  spectrum and states of the harmonic oscillator are solved.   We give a summary of our results  in section \ref{sec3}.

\section{Position-dependent NC quantum mechanics}\label{sec1}
This section addresses the construction of a position-dependent NC star product which is  induced by the deformation (\ref{deformation}). We start with the following definition.
\begin{definition}[Twisted Moyal algebra]
Consider the set  $E=\{(x^i,p^i), i=1,2\}$ and $\mathbb{C}[[x^{1},
x^{2},p^1,p^2]],$ the free algebra generated by $E$. Let $\mathcal{I}$ be
the ideal of $\mathbb{C}[[x^{1}, x^{2},p^1,p^2]],$ generated by the
elements 
$
[x^i,x^j]_\star-i\theta^{ij}(x),\quad [x^i,p^j]_\star-i\hbar_{eff}\delta^{ij},\quad
[p^i,p^j]_\star-i\bar\theta^{ij},
$
where $\theta^{ij}(x)$ is skew symmetric tensor depending on space  coordinates and $\bar\theta^{ij}$ a constant skew symmetric tensor. 
The
twisted Moyal algebra $\mathcal{M}_{\theta\bar\theta\hbar_{eff}}$ is the quotient
$\mathbb{C}[[x^{1}, x^{2},p^1,p^2]]/\mathcal{I}$. Each element in
$\mathcal{M}_{\theta\bar\theta\hbar_{eff}}$ is a formal power series in the $(x^i,p^j)$'s
for which the following  relations hold:
\bea\label{defstar}
[x^i,x^j]_\star=i\theta^{ij}(x),\quad [x^i,p^j]_\star=i\hbar_{eff}\delta^{ij},\quad
[p^i,p^j]_\star=i\bar\theta^{ij}.
\eea
\end{definition}
 The  Moyal algebra can be
 also defined as  the linear space of smooth and rapidly
decreasing functions equipped with the  NC star product given in  the form $f\star g={\bf m}\big[(\star_\theta \star_{\hbar_{eff}}\star_{\bar\theta})(f\otimes g)\big] $, such that
\bea\label{star1}
&&f\star_\theta g={\bf m}\Big[\exp\Big(\frac{i}{2}\theta^{ij}(x)\partial_{x^i}
\otimes\partial_{x^j}\Big)f\otimes g\Big]\\
&&f\star_{\hbar_{eff}} g={\bf m}\Big[\exp\Big(\frac{i}{2}\hbar_{eff}\delta^{ij}(\partial_{x^i}
\otimes\partial_{p^j}-\partial_{p^i}
\otimes\partial_{x^j})\Big)f\otimes g\Big]\\
&&f\star_{\bar\theta}g ={\bf m}\Big[\exp\Big(\frac{i}{2}\bar\theta^{ij}\partial_{p^i}
\otimes\partial_{p^j}\Big)f\otimes g\Big].
\eea
%Remark that with measure $d\mu=\sqrt{F(x)}d^Dx$ we get the trace relation
%\bea
%\int \sqrt{F(x)}d^Dx\, (f\star g)(x)=\int \sqrt{F(x)}d^Dx\, (g\star f)(x)=\int \sqrt{F(x)}d^Dx\, f(x)\cdot g(x).
%\eea
% In a parallel direction we can construct the new star product by
%\bea
%f\star' g=D^{-1}(Df\star Dg)
%\eea
For $D=2$, we set  $x^i=(x^1,x^2)$, $p^i=(p^1,p^2)$ with $(x^i,p^i)\in\mathbb{R}^4$ and we will restrict the NC structure tensors to the following:
\bea
\theta^{ij}(x)=\theta e(x)\left(\br{cc}
0&1\\
-1&0
\er\right)=\theta e(x)\epsilon^{ij},\quad  \bar\theta^{ij}=\bar\theta\left(\br{cc}
0&1\\
-1&0
\er\right)=\bar\theta\epsilon^{ij}.
\eea
 For   $f\in C^\infty(\mathbb{R}^4)$, the following relations are satisfied
\bea\label{detail}
x^1\star f=x^1 f+\frac{i\theta e}{2}\partial_{x^2}f+\frac{i\hbar_{eff}}{2}\partial_{p^1}f,\qquad p^1\star f=p^1 f+\frac{i\bar\theta}{2}\partial_{p^2}f-\frac{i\hbar_{eff}}{2}\partial_{x^1}f,\\
\label{detail1}x^2\star f=x^2 f-\frac{i\theta e}{2}\partial_{x^1}f+\frac{i\hbar_{eff}}{2}\partial_{p^2}f,\qquad
p^2\star f=p^2 f-\frac{i\bar\theta}{2}\partial_{p^1}f-\frac{i\hbar_{eff}}{2}\partial_{x^2}f.
\eea
The commutation relation (\ref{defstar}) can be deduced from (\ref{detail}) and  (\ref{detail1}).
%Noting  that  in the case where $e=1$, the effective Planck constant can be expressed as 
%$
% \hbar_{eff}=\hbar\Big(1-\frac{\theta\bar\theta}{8\hbar^2}\Big),
%$
% and we recover the ordinary Moyal algebra on the  phase space study in more literatures. 

We can now introduce a model on that twisted NC space. Let us  consider  the NC harmonic oscillator  described by the Hamiltonian
\bea\label{hamilton}
H=\frac{1}{2}\Big[(p^1)^2+(p^2)^2+(x^1)^2+(x^2)^2
\Big].
\eea
In  (\ref{hamilton}), the mass parameter  and the oscillator constant  are taken to be $1$. The Hamiltonian (\ref{hamilton}) is invariant under the  phase space rotation.  We now address the problem we want to solve. 

\medskip 

We will solve  the eigenvalue problem associated with \eqref{hamilton} in the NC phase space
 \bea\label{eig}
H\star\psi=E\,\psi.
\eea 
A way to solve the  eigenvalue problem of a quantum Hamiltonian is its factorization. In the following, will introduce a particular type of factorization.
The eigenvalue problem \eqref{eig}  
can be split  into two equations given by
\bea\label{MO}
H_R\star\psi=E\,\psi,\quad\mbox{and}\quad H_{Im}\star \psi=0
\eea
where, by expansion of the twisted star product, one should obtain the real and imaginary part corresponding to \eqref{eig} as:
\bea\label{HO1}
H_R&=&\frac{1}{2}\Big[(p^1)^2+(p^2)^2+(x^1)^2+(x^2)^2+
+\frac{\bar\theta \hbar_{eff}}{2}(\partial_{p^2}\partial_{x^1}-\partial_{p^1}
\partial_{x^2})
\cr
&-&\frac{\theta \hbar_{eff}e}{2}(\partial_{x^2}\partial_{p^1}-\partial_{x^1}
\partial_{p^2})
-\frac{\hbar_{eff}^2}{4}(\partial_{p^1}^2+\partial_{p^2}^2+\partial_{x^1}^2
+\partial_{x^2}^2)\cr
&-&\frac{\bar\theta^2}{4}(\partial_{p^1}^2+\partial_{p^2}^2)-\frac{\theta^2 e}{4}(\partial_{x^1}^2+\partial_{x^2}^2)-\frac{\theta e}{4}(\omega_2\partial_{x^2}+\omega_1\partial_{x^1})\Big]
\eea
and
\beq
\label{HO2}
H_{Im}=\bar{\theta}(p^1\partial_{p^2}-p^2\partial_{p^1})+\theta
e(x^1\partial_{x^2}-x^2\partial_{x^1})+\hbar_{eff}(x^1\partial_{p^1}+x^2\partial_{p^2}-p^1\partial_{x^1}-p^2\partial_{x^2}).
\eeq
Note that the eigenvalue equation (\ref{eig}) can be also written as $\psi\star H=E\psi$, due to the fact that $f\star g=\overline{g\star f}$. Equations (\ref{MO}) with the Hamiltonians (\ref{HO1}) and (\ref{HO2}) are nonlinear. However after putting a restriction on the type  of noncommutativity that we use, we will provide solution to these equations.

\medskip  

{\bf Solution -} Consider $e(x)=1+\omega_1 x^1+\omega_2 x^2$. Let us assume that $\omega_1\neq 0$ or $\omega_2\neq 0$.  Consider the
 transformation $\mathcal T$ mapping coordinates $(x,p)$ to the new variables  $(\widetilde x, \widetilde p)$ given by 
\bea\label{transsuper}
\mathcal{T}:\left\{\begin{array}{c}
x^1=\theta\omega_1 e^{\widetilde x^1}-\theta\omega_2\widetilde x^2-
\frac{\omega_1}{\omega_1^2+\omega_2^2}\\ 
x^2=\theta\omega_2 e^{\widetilde x^1}+\theta\omega_1\widetilde x^2
-
\frac{\omega_2}{\omega_1^2+\omega_2^2}\\
 p^1=\widetilde p^1\\
  p^2=\widetilde p^2
\end{array}\right.
\eea
For $\theta>0,$ the transformation $\mathcal T$ is invertible in the positive domain of the plane  $(x^1, x^2)$ given by relation
\beq
e(x)=1+\omega_1 x^1+\omega_2 x^2>0.
\eeq
The inverse transformation $\mathcal{T}^{-1}$ is    given by
\bea\label{tr}
\mathcal{T}^{-1}:\left\{\begin{array}{c}
 \widetilde x^1=\ln\Big(\frac{e(x)}{\theta(\omega_1^2+\omega_2^{2})}\Big)\\
\widetilde x^2=\frac{-\omega_2x^1+\omega_1x^2}{\theta(\omega_1^2+\omega_2^{2})}\\ \widetilde p^1=p^1\\ \widetilde p^2=p^2
\end{array}\right.
\eea
Let us immediately remark that the transformation 
\eqref{tr} break the ordinary limit of the theory i.e. the limit $\theta\rightarrow 0$ cannot be taken into account. To recover this inconvenience we can use the renormalization procedures, which will be addressed in forthcoming work.

Under $\mathcal{T}$, the algebra $\mathcal{M}_{\theta\bar\theta\hbar_{eff}}$ is transformed  as  $\widetilde{\mathcal{M}}_{\theta\bar\theta\hbar_{eff}}=\mathcal{ T}[\mathcal{M}_{\theta\bar\theta\hbar_{eff}}]$.  The non-vanishing commutation relations   satisfied by the new variables are given by the 
 following:
\begin{eqnarray}\label{comnew}
&&[\widetilde x^1,\widetilde x^2]_\star=i\theta\sqrt{\omega_1^2+\omega_2^{2}}
=i\gamma,\quad [\widetilde x^1,\widetilde p^1]_\star=i\hbar_{eff}\omega_1 e^{-1}=i\hbar_{1}(x),\\ \label{comnew1}&&[\widetilde x^2,\widetilde p^2]_\star=i\frac{\omega_1\hbar_{eff}}{\theta(\omega_1^2+\omega_2^{2})}=i\hbar_{2},\quad [\widetilde p^1,\widetilde p^2]_\star=i\bar\theta.
\end{eqnarray}

The ordinary recipes that are known for diagonalizing the algebra do not work in this instance because of the presence of the function $e(x)$. However,  restricting to the case
 $\omega_1=0$ and $\omega_2\neq 0,$ we have  
\bea\label{ntran}
x^1=-\gamma\widetilde x^2,\quad x^2=\gamma e^{\widetilde x^1}
-
\frac{1}{\omega_2},\quad p^1=\widetilde p^1,\quad  p^2=\widetilde p^2,\quad \gamma=\theta\omega_2.
\eea 
Therefore, for simplicity by setting $\omega_1=0$ and $\omega_2=1=\gamma$,
we understand that the transformation $\mathcal{T}$
simply induces a rotation in the plane $(x^1,x^2) \to (x^2, -x^1)$  
followed by a logarithmic scale transformation $(x^1,x^2)\to ( \ln [x^1+1],x^2)$. Note that such a transformation cannot 
be defined in the case of the Moyal plane determined by the
limiting situation $\omega_1=\omega_2=0$. Furthermore, it can be noticed that the restriction \eqref{ntran} clearly breaks the symmetry between the coordinates $x^1$ and $x^2$.  In any case, the following analysis finds an analog when we consider $\omega_2=0,\,\,\omega_1\neq 0$.

We obtain the final    commutation relations 
\bea\label{bommm}
 [\widetilde x^1,\widetilde p^1]_\star=0,\quad  [\widetilde x^2,\widetilde p^2]_\star=0,\quad  [\widetilde x^1,\widetilde x^2]_\star=i\gamma,\quad 
 [\widetilde p^1,\widetilde p^2]_\star=i\bar\theta.
\eea
As a consequence, the algebra  $\widetilde{\mathcal{M}}_{\theta\bar\theta\hbar_{eff}}$  splits into two sectors $\widetilde{\mathcal{M}}_{\theta}$ and $\widetilde{\mathcal{M}}_{\bar{\theta}}$ such that
\bea
\widetilde{\mathcal{M}}_{\theta}\otimes \widetilde{\mathcal{M}}_{\bar\theta}\equiv  \widetilde{\mathcal{M}}_{\theta\bar\theta\hbar_{eff}},
\eea
where the algebras $\widetilde{\mathcal{M}}_{\theta}$ and $\widetilde{\mathcal{M}}_{\bar\theta}$ are each of the Moyal-type 
defined such that
$
\widetilde{\mathcal{M}}_{\theta}=\mathbb{C}[[\widetilde x^{1}, \widetilde x^{2}]]/\mathcal{I}_1$ and 
$\widetilde{\mathcal{M}}_{\bar\theta}=\mathbb{C}[[\widetilde p^{1}, \widetilde p^{2}]]/\mathcal{I}_2,
$
where $\mathcal{I}_1$ and $\mathcal{I}_2$ are, respectively,
the ideal of $\mathbb{C}[[\widetilde x^{1}, \widetilde x^{2}]],$ generated by the
elements $[\widetilde x^1,\widetilde x^2]_\star-i\gamma$, and  the ideal of $\mathbb{C}[[\widetilde p^{1}, \widetilde p^{2}]],$ generated by the
elements $[\widetilde p^1,\widetilde p^2]_\star-i\bar\theta$.
In short, $\widetilde{\mathcal{M}}_{\theta}\otimes \widetilde{\mathcal{M}}_{\bar\theta}$ defines a standard 4 dimensional
Moyal space $[y^{\alpha}, y^{\beta} ]= \Theta^{\alpha\beta}$,
$\alpha,\beta=1,2,3,4$,  with tensor structure 
\beq
\Theta:= \left(\begin{array}{cc}
\gamma J &0 \\
0 & \theta J 
\end{array}\right) \qquad  
J : = \left(\begin{array}{cc}
0 &1 \\
-1 & 0 
\end{array}\right)
\eeq
where $y^{1,2}=\widetilde{x}^{1,2}$
and $y^{3,4}=\widetilde{p}^{1,2}$.

For simplicity, we  set 
$\gamma=1$, $\omega_2=2$ and $\bar\theta =1$.
Using   (\ref{ntran}),
the Hamiltonian (\ref{hamilton}) takes the form
\bea
H=\frac{1}{2}\Big[\gamma^2(\widetilde x^2)^2+\gamma^2 e^{2\widetilde x^1}-\frac{2\gamma}{\omega_2}e^{\widetilde x^1}+\frac{1}{\omega_2^2}+(\widetilde p^1)^2+(\widetilde p^2)^2\Big]=H_1(\widetilde x^1,\widetilde x^2)+H_2(\widetilde p^1,\widetilde p^2)
\eea
where
\bea
H_1(\widetilde x^1,\widetilde x^2)=\frac{1}{2}\Big[(\widetilde x^2)^2+ e^{2\widetilde x^1}-e^{\widetilde x^1}+\frac14\Big],\quad
H_2(\widetilde p^1,\widetilde p^2)=\frac{1}{2}\Big[(\widetilde p^1)^2+(\widetilde p^2)^2\Big].
\eea
Now using the commutation relations \eqref{bommm}, we get
\bea
[H_1(\widetilde x^1,\widetilde x^2),H_2(\widetilde p^1,\widetilde p^2)]_\star=0.
\eea
It appears clear that the star product $\star=\star_\theta\star_{\hbar_{eff}}\star_{\bar\theta} $ gets mapped 
as
\beq
\star \longrightarrow \mathcal T (\star) = \star_1\star_2 
\eeq
with 
\bea
\star_1={\bf m}\Big[\exp\Big(\frac{i}{2}(\partial_{\widetilde x^1}\otimes\partial_{\widetilde x^2}-\partial_{\widetilde x^2}\otimes\partial_{\widetilde x^1})\Big)\Big],\quad
\star_2={\bf m}\Big[\exp\Big(\frac{i}{2}(\partial_{\widetilde p^1}\otimes\partial_{\widetilde p^2}-\partial_{\widetilde p^2}\otimes\partial_{\widetilde p^1})\Big)\Big].
\eea
Then, the  new coordinate and momentum operators can be described by the following relations
\bea\label{prop1}
\widetilde x^1\star_1=\widetilde x^1+\frac{i}{2}\partial_{\widetilde x^2},\quad 
\widetilde x^2\star_1=\widetilde x^2-\frac{i}{2}\partial_{\widetilde x^1},\quad  \widetilde p^1\star_2=\widetilde p^1+\frac{i}{2}\partial_{\widetilde p^2},\quad
 \widetilde p^2\star_2=\widetilde p^2-\frac{i}{2}\partial_{\widetilde p^1}.
\eea

The initial Hamiltonian has been factorized into two commuting sectors. 
We can first study the spectrum of Hamiltonian $H_1(\widetilde x^1,\widetilde x^2)$ so called supersymmetric  Liouville Hamiltonian.
Using (\ref{prop1}), the Hamiltonian in this sector  takes the form
\bea
H_1(\widetilde x^1,\widetilde x^2)\star_1&=&\frac{1}{2}\Big[(\widetilde x^2)^2-i\widetilde x^2\partial_{\widetilde x^1}-\frac{1}{4}\partial^2_{\widetilde x^1}+e^{2\widetilde x^1}(\cos\partial_{\widetilde x^2}+i\sin\partial_{\widetilde x^2})\cr
&-&e^{\widetilde x^1}(\cos\frac{1}{2}\partial_{\widetilde x^2}+i\sin\frac{1}{2}\partial_{\widetilde x^2})+\frac{1}{4}\Big].
\eea
For a real $E_1$ and  a wave function $\psi_{1,E_1},$ the   eigenvalue problem 
$
H_1(\widetilde x^1,\widetilde x^2)\star_1 \psi_{1,E_1}=E_1\psi_{1,E_1}$  can be re-expressed into two parts: the real part is given by
\bea\label{real}
\Big((\widetilde x^2)^2-\frac{1}{4}\partial^2_{\widetilde x^1}+e^{2\widetilde x^1}\cos\partial_{\widetilde x^2}
-e^{\widetilde x^1}\cos\frac{1}{2}\partial_{\widetilde x^2}+\frac{1}{4}-2E_1\Big)\psi_{1,E_1}=0
\eea
whereas the imaginary part expresses as
\bea\label{im}
\Big(\widetilde x^2\partial_{\widetilde x^1}-e^{2\widetilde x^1}\sin\partial_{\widetilde x^2}
+e^{\widetilde x^1}\sin\frac{1}{2}\partial_{\widetilde x^2}\Big)\psi_{1,E_1}=0.
\eea

To solve consistently the equations (\ref{real}) and  (\ref{im}), we will use a fact about the Taylor expansion of  an arbitrary function $\psi(x)$,
for the small values of parameter $\epsilon$ as
\bea\label{at1}
\psi(x+\epsilon)=\psi(x)+\epsilon\partial_x\psi(x)+
\frac{1}{2}\epsilon^2\partial^2_x\psi(x)+
\cdots=e^{\epsilon\partial_x}\psi(x)\\
\label{at2}\psi(x-\epsilon)=\psi(x)-\epsilon\partial_x\psi(x)+
\frac{1}{2}\epsilon^2\partial^2_x\psi(x)+
\cdots=e^{-\epsilon\partial_x}\psi(x).
\eea
Then summing (\ref{at1}) and (\ref{at2}), we get
\bea
&&\frac{1}{2}\Big(\psi(x+\epsilon)+\psi(x-\epsilon)\Big)=\cosh \epsilon\partial_x\,\psi(x),\\
&& \frac{1}{2}\Big(\psi(x+\epsilon)-\psi(x-\epsilon)\Big)=\sinh \epsilon\partial_x \,\psi(x).
\eea
We restrict to the case where $\epsilon=i$  and $\epsilon=\frac{i}{2}$.   % and this yields
%\bea
%&&\frac{1}{2}\Big(\psi(x+i)+\psi(x-i)\Big)=\cos \partial_x\,\psi(x),\\ &&\frac{1}{2}\Big(\psi(x+i)-\psi(x-i)\Big)=i\sin \partial_x\,\psi(x)\\
%&&\frac{1}{2}\Big(\psi(x+\frac{i}{2})+\psi(x-\frac{i}{2})\Big)=\cos \frac{1}{2}\partial_x\,\psi(x),\\ 
%&&\frac{1}{2}\Big(\psi(x+\frac{i}{2})-\psi(x-\frac{i}{2})\Big)=i\sin \frac{1}{2}\partial_x\,\psi(x).
%\eea
Then follow from the identities
\bea
\Big(\sin\partial_{\widetilde x^2}\Big)\psi(\widetilde x^1,\widetilde x^2)=\frac{1}{2i}\Big(\psi(\widetilde x^1,\widetilde x^2+i)-\psi(\widetilde x^1,\widetilde x^2-i)\Big)\\
\Big(\cos\partial_{\widetilde x^2}\Big)\psi(\widetilde x^1,\widetilde x^2)=\frac{1}{2}\Big(\psi(\widetilde x^1,\widetilde x^2+i)+\psi(\widetilde x^1,\widetilde x^2-i)\Big)
\eea
and
\bea\label{idnew}
\Big(\sin\frac{1}{2}\partial_{\widetilde x^2}\Big)\psi(\widetilde x^1,\widetilde x^2)=\frac{1}{2i}\Big(\psi(\widetilde x^1,\widetilde x^2+i/2)-\psi(\widetilde x^1,\widetilde x^2-i/2)\Big)\\
\Big(\cos\frac{1}{2}\partial_{\widetilde x^2}\Big)\psi(\widetilde x^1,\widetilde x^2)=\frac{1}{2}\Big(\psi(\widetilde x^1,\widetilde x^2+i/2)+\psi(\widetilde x^1,\widetilde x^2-i/2)\Big).
\eea
The  equation (\ref{im}) can be simply written as
\bea\label{aasin}
\Big(\widetilde x^2\partial_{\widetilde x^1}\Big)\psi_{1,E}(\widetilde x^1,\widetilde x^2)&=&\frac{e^{2\widetilde x^1}}{2i}\Big(\psi_{1,E}(\widetilde x^1,\widetilde x^2+i)-\psi_{1,E}(\widetilde x^1,\widetilde x^2-i)\Big)\cr
&-&\frac{e^{\widetilde x^1}}{2i}\Big(\psi(\widetilde x^1,\widetilde x^2+i/2)-\psi(\widetilde x^1,\widetilde x^2-i/2)\Big).
\eea
Using the above relations,
we then  get the new equation corresponding to \eqref{real} as
\bea\label{realnew}
&&\Big((\widetilde x^2)^2+\frac{1}{4}-2E_1\Big)\psi_{1,E_1}(\widetilde x^1,\widetilde x^2)    -\frac{1}{4}\Big[\frac{e^{2\widetilde x^1}}{i\widetilde x^2}\Big(\psi_{1,E}(\widetilde x^1,\widetilde x^2+i)-\psi_{1,E}(\widetilde x^1,\widetilde x^2-i)\Big)\cr
&&-\frac{e^{\widetilde x^1}}{2i\widetilde x^2}\Big(\psi(\widetilde x^1,\widetilde x^2+i/2)-\psi(\widetilde x^1,\widetilde x^2-i/2)\Big)
-
\frac{e^{4\widetilde x^1}}{4(\widetilde x^2)^2}\Big(\psi(\widetilde x^1,\widetilde x^2+2i)-\psi(\widetilde x^1,\widetilde x^2)\Big)\cr
&&+
\frac{e^{2\widetilde x^1}}{4(\widetilde x^2)^2}\Big(\psi(\widetilde x^1,\widetilde x^2+i)-\psi(\widetilde x^1,\widetilde x^2)+2\psi(\widetilde x^1,\widetilde x^2-i)\Big)\Big]\cr
&&-\frac{e^{2\widetilde x^1}}{2}\Big(\psi(\widetilde x^1,\widetilde x^2+i)+\psi(\widetilde x^1,\widetilde x^2-i)\Big)
+\frac{e^{\widetilde x^1}}{2}\Big(\psi(\widetilde x^1,\widetilde x^2+i/2)+\psi(\widetilde x^1,\widetilde x^2-i/2)\Big)\Big]=0.\nonumber\\
\eea
In \cite{Curtright:1997me}, the recursive properties of the Meijer G-function can be used to compute
the eigenvectors $\psi_{1,E_1}(\widetilde x^1)$ as
\bea\label{toto}
\psi_{1,E_1}(\widetilde x^1)=\Big(\frac{1}{4\pi^2\sqrt{E_1}}e^{\widetilde x^1}\cosh(\pi\sqrt{E_1})\Big)^{1/2}\Big(K_{\frac{1}{2}-i\sqrt{E_1}}(e^{\widetilde x^1})+K_{\frac{1}{2}+i\sqrt{E_1}}(e^{\widetilde x^1})\Big),\,\,E_1\geq 0,
\eea
where $K$ are Kelvin (modified Bessel) functions. 

 The second eigenvalue problem 
$
H_2(\widetilde p^1,\widetilde p^2)\star_2 \psi_{2,E_2}=E_2\psi_{2,E_2}
$ is well known as the simple quantum harmonic oscillator problem. We write
\bea
\psi_{2,E_2}(\widetilde p^1)&=&\Big(\frac{1}{\pi}\Big)^{1/4}\frac{1}{2^nn!}H_n(\widetilde p^1) e^{-(\widetilde p^1)^2/2}
\eea
where $H_n$ stand for the Hermite polynomial and the oscillator energy as $E_2=n+\frac{1}{2}$.
Finally the solution of Hamiltonian (\ref{hamilton}) is then
\bea\label{solution}
\psi_{E}(\widetilde x^1,\widetilde p^1)=\psi_{1,E_1}(\widetilde x^1)\otimes\psi_{2,E_2}(\widetilde p^1),\quad E=E_1+E_2.
\eea
We conclude that the spectrum of the Hamiltonian $H$ is composed by two sectors: A continuum part in the sector  $H_1$ and a discrete one in the sector $H_2$.

\section{Conclusion}\label{sec3}
In this work, following our previous approach \cite{Hounkonnou:2010yk} and results based on  \cite{Curtright:1997me}, we have found the eigenvalues and eigenvectors of the harmonic oscillator in the twisted Moyal space
with function structure $e(x)=1+\omega_2 x^2$. We have introduced a particular transformation which has allowed us to split the total twisted Moyal algebra into two parts in which the Hamiltonian was written  in two commuting pieces.
 
Let us remark that the solution \eqref{solution} exhibits the lack of commutative limit $\theta\rightarrow 0$. This inconvenience  is due to the form of the scale transformation $\mathcal{T}$. Therefore the solution obtained in relation \eqref{toto} need to be renormalized. Note also that
it is  a more difficult problem to find a solution for the harmonic oscillator in the more symmetric case of $e(x)=1+\omega_\mu x^\mu.$ This two 	
tangles deserves to be investigated.

\section*{Acknowledgements}
The author would like to thank Joseph Ben Geloun for  useful comments which have improved this work.
This research was supported in part by Perimeter Institute for Theoretical Physics. Research at Perimeter Institute is supported by the Government of Canada through Industry Canada and by the Province of Ontario through the Ministry of Research and Innovation.

\end{document}